\documentclass[aps,pra,reprint,amsmath,amssymb,showpacs,floatfix]{revtex4-1}

\usepackage{graphicx,units}

\newcommand{\ket}[1]{\ensuremath{\vert{#1\rangle}}} 
\newcommand{\bra}[1]{\ensuremath{{\langle #1}\vert}}
\newcommand{\braket}[2]{\ensuremath{{\langle #1}\vert{#2 \rangle}}}
\newcommand{\ketbra}[2]{\ensuremath{|{#1 \rangle}{\langle #2}|}}
\newcommand{\op}[1]{\hat{#1}}
\newcommand{\D}{\text{d}}
\newcommand{\I}{\text{i}}
\newcommand{\E}{\text{e}}
\providecommand{\abs}[1]{\left\lvert#1\right\rvert}
\newcommand{\bvec}[1]{\ensuremath{\mathbf{#1}}}

\newcommand{\buvec}[1]{\ensuremath{\mathbf{\hat{#1}}}}

\newcommand{\bopvec}[1]{\ensuremath{\mathbf{\op{#1}}}}
\newcommand{\bopvecgr}[1]{\ensuremath{\mathbf{\op{\boldsymbol #1}}}}
\newcommand{\boldnabla}{\mbox{\boldmath$\nabla$}}

\usepackage[colorlinks,citecolor=blue,urlcolor=blue,linkcolor=blue,hyperindex,breaklinks]{hyperref}
\begin{document}

\title{Protective measurement of open quantum systems}

\author{Maximilian Schlosshauer}

\affiliation{Department of Physics, University of Portland, 5000 North Willamette Boulevard, Portland, Oregon 97203, USA}

\begin{abstract}
We study protective quantum measurements in the presence of an environment and decoherence. We consider the model of a protectively measured qubit that also interacts with a spin environment during the measurement. We investigate how the coupling to the environment affects the two characteristic properties of a protective measurement, namely, (i) the ability to leave the state of the system approximately unchanged and (ii) the transfer of information about expectation values to the apparatus pointer. We find that even when the interaction with the environment is weak enough not to lead to appreciable decoherence of the initial qubit state, it causes a significant broadening of the probability distribution for the position of the apparatus pointer at the conclusion of the measurement. This washing out of the pointer position crucially diminishes the accuracy with which the desired expectation values can be measured from a readout of the pointer. We additionally show that even when the coupling to the environment is chosen such that the state of the system is immune to decoherence, the environment may still detrimentally affect the pointer readout.\\[.2cm]
Journal reference: \emph{Phys.\ Rev.\ A} {\bf 101}, 012108 (2020),  \href{https://doi.org/10.1103/PhysRevA.101.012108}{\texttt{10.1103/PhysRevA.101.012108}}
\end{abstract}

\maketitle

\section{Introduction}

Quantum measurements in which an apparatus is weakly coupled to a quantum system play an important role in the investigation of quantum phenomena \cite{Dressel:2014:uu,Gao:2014:cu}. Two main categories of such weak measurements have been studied:  instantaneous weak measurements \cite{Aharonov:1988:mz,Dressel:2014:uu} and protective measurements \cite{Aharonov:1993:qa,Aharonov:1993:jm,Dass:1999:az,Vaidman:2009:po,Gao:2014:cu,Genovese:2017:zz,Qureshi:2015:jj}. In instantaneous weak measurements (usually simply called weak measurements), the apparatus interacts with the system only momentarily, followed by postselection. The shift of the apparatus pointer then encodes the weak value \cite{Aharonov:1988:mz,Dressel:2014:uu} of a system observable $\op{A}$ for given pre- and postselected states \cite{Aharonov:1988:mz,Duck:1989:uu,Dressel:2014:uu}. By contrast, in protective measurements \cite{Aharonov:1993:qa,Aharonov:1993:jm,Dass:1999:az,Vaidman:2009:po,Gao:2014:cu,Genovese:2017:zz,Qureshi:2015:jj} the apparatus is coupled to the system not instantaneously but for a time $T$ much longer than the intrinsic timescale of the system. If the system starts out in an eigenstate of its Hamiltonian, then this state remains approximately unchanged during the measurement while the apparatus pointer is shifted by an amount proportional to the expectation value of $\op{A}$ in the initial state \footnote{A different version of protective measurements, which we shall not consider here, uses repeated projective measurements to protect the state \cite{Aharonov:1993:jm,Piacentini:2017:oo}.}. Applications of protective measurements include the direct measurement of the quantum state of a single system \cite{Aharonov:1993:qa,Aharonov:1993:jm,Aharonov:1996:fp,Dass:1999:az,Vaidman:2009:po,Auletta:2014:yy,Diosi:2014:yy,Aharonov:2014:yy,Schlosshauer:2016:uu}, studies of particle trajectories \cite{Aharonov:1996:ii,Aharonov:1999:uu}, determination of stationary states \cite{Diosi:2014:yy}, translation of ergodicity into the quantum realm \cite{Aharonov:2014:yy}, fundamental investigations of quantum measurement \cite{Aharonov:1993:qa,Aharonov:1993:jm,Aharonov:1996:fp, Alter:1996:oo,Alter:1997:oo,Dass:1999:az,Gao:2014:cu}, and the description of two-state thermal ensembles \cite{Aharonov:2014:yy}. An experimental realization of a protective measurement using photons has been reported in Ref.~\cite{Piacentini:2017:oo}.

In general, any realistic quantum system is open and consequently the state of the system is subject to decoherence due to interactions with the environment \cite{Zurek:1982:tv,Zurek:2002:ii,Schlosshauer:2019:qd}. For instantaneous weak measurements, the decoherence acting between the preselection and the start of the measurement, and again between the end of measurement and the postselection, will influence the measured weak value \cite{Shikano:2010:zz,Abe:2015:yy}. However, because the measurement is instantaneous, the measurement and decoherence interactions can be treated independently and the effect of decoherence can be minimized by performing the pre- and postselection close to the time of measurement. The situation is very different and more acute in a protective measurement, since here the system must be coupled to the apparatus for a long time, during which the system will also be subject to interactions with its environment. Therefore, the dynamics will be governed simultaneously by the measurement and decoherence interactions, and the environment can substantially affect the protective measurement in two ways. First, because decoherence will in general change the quantum state, it decreases the rate of success of the measurement (since an ideal protective measurement desires the state of the system to remain unaltered). Second, environmental interactions can influence the state of the apparatus pointer at the conclusion of the measurement, thereby diminishing the ability of the pointer to accurately reveal the expectation value of the chosen observable. 

In this paper, we consider a generic model for the protective measurement of a qubit and extend it by adding to it the interaction of the qubit with an environment of other two-level systems. We study, for different strengths of the environmental interaction, the resulting evolution of the state of the qubit and the apparatus pointer. 

This paper is organized as follows. In Sec.~\ref{sec:model}, we describe our model (developing it, for concreteness, in the context of a Stern--Gerlach measurement setup) and solve for its dynamics. We then investigate the influence of the environment on the initial state of the system (Sec.~\ref{sec:infl-decoh-init}) and on the shift of the apparatus pointer (Sec.~\ref{sec:infl-decoh-point}). In Sec.~\ref{sec:envir-effects-with}, we discuss how the environment can negatively affect the apparatus pointer even when it does not change the state of the system. In Sec.~\ref{sec:stern-gerl-exper}, we describe a scheme for an experimental test of our model using a setup of the Stern--Gerlach type. In Sec.~\ref{sec:gener-appl}, we show that our model and results are general in the sense that they apply  beyond the Stern--Gerlach scenario to any protective measurement of a qubit system in contact with a spin environment. We discuss our results in Sec.~\ref{sec:discussion}.

\section{\label{sec:model}Model and dynamics}

\subsection{Protective measurement}

A general protective measurement of a system $S$ by an apparatus $A$ can be described by the Hamiltonian
\begin{equation}\label{eq:3aaa}
\op{H}(t) = \op{H}_S+\op{H}_m(t)= \op{H}_S+ \kappa(t) \op{O}_S \otimes \op{K}_A,
\end{equation}
where $\op{H}_S$ is the self-Hamiltonian of the system and $\op{H}_m$ represents the measurement interaction between system and apparatus. $\op{O}_S$ is an arbitrary observable of the system, and $\op{K}_A$ is an operator that generates the shift of the apparatus pointer. The function $\kappa(t)$ is a time-dependent coupling strength, which we take to be proportional to $1/T$ during the duration $t \in [0,T]$ of the measurement, and equal to zero otherwise (more complicated time dependencies may also be considered \cite{Schlosshauer:2014:pm}). The measurement is weak in the sense that $T$ is chosen sufficiently long such that $\op{H}_S$ dominates. If the system starts out in an eigenstate $\ket{\psi}$ of $\op{H}_S$, the probability of transitioning to a different state at the conclusion of the measurement interaction can be made arbitrarily small by increasing $T$ (and thus making the measurement interaction longer and weaker), and the apparatus pointer shifts by an amount proportional to the expectation value $\bra{\psi}\op{O}_S\ket{\psi}$. In this way, the state is effectively protected by $\op{H}_S$ and one may, for example, reconstruct the quantum state of a single system from protective measurements of a complete set of observables \cite{Aharonov:1993:qa,Vaidman:2009:po,Gao:2014:cu,Schlosshauer:2016:uu}. 

We now focus on the case of a qubit system (with self-Hamiltonian $\op{H}_S=\frac{1}{2}\hbar \omega_0 \op{\sigma}_z$) on which a generic qubit observable $\op{O}_S=\bopvecgr{\sigma} \cdot \buvec{m}$ is protectively measured. Then the Hamiltonian~\eqref{eq:3aaa} becomes
\begin{equation}\label{eq:7}
\op{H}(t) = \frac{1}{2}\hbar \omega_0 \op{\sigma}_z + \kappa (t) (\bopvecgr{\sigma} \cdot \buvec{m}) \otimes \op{K}_A.
\end{equation}
For concreteness, and to make contact with models of protective measurement studied previously \cite{Aharonov:1993:jm,Dass:1999:az,Schlosshauer:2015:uu}, we shall consider a realization of the Hamiltonian~\eqref{eq:7} in a setting of the Stern--Gerlach type, describing a spin-$\frac{1}{2}$ particle subject to magnetic fields. We stress, however, that our calculations and results are not tied to this particular realization. They are generic in the sense that they apply to the protective measurement of any qubit system by an apparatus pointer as described by Eq.~\eqref{eq:7}. We discuss this generality and applications beyond the Stern--Gerlach setting in Sec.~\ref{sec:gener-appl} below.

In the scenario of the Stern--Gerlach type, $\op{H}_S$ corresponds to a uniform protection field $\bvec{B}_0$ in the $+z$ direction,
\begin{equation}\label{eq:vshvbjfdjhvs}
\op{H}_S = -  \mu \bopvecgr{\sigma} \cdot \bvec{B}_0 = -\mu B_0 \op{\sigma}_z,
\end{equation}
where $\mu$ denotes the magnetic moment of the particle. The eigenstates of $\op{H}_S$ are the eigenstates $\ket{0}$ and $\ket{1}$ of $\op{\sigma}_z$, with eigenvalues $E_\pm=\mp \mu B_0$ and corresponding transition frequency $\omega_0 = 2\mu B_0/\hbar$. During the measurement interval $[0,T]$, the particle additionally experiences an inhomogeneous measurement field given by \footnote{Because the field given by Eq.~\eqref{eq:measfield} has nonzero divergence, it violates Maxwell's equations and therefore cannot represent a physical magnetic field, as already noted in Ref.~\cite{Anandan:1993:uu}. However, one can easily construct a suitable divergence-free inhomogeneous field and show that it produces the same pointer shift and state disturbance \cite{Anandan:1993:uu,Schlosshauer:2015:uu}. Therefore, for our purposes it suffices to consider, without loss of generality, the field given by Eq.~\eqref{eq:measfield}.}
\begin{equation}\label{eq:measfield}
\bvec{B}_m(\bvec{x}) = \frac{1}{T} \beta q  \buvec{m},
\end{equation}
where $q$ is the position coordinate in the field direction given by the unit vector $\buvec{m}$. We specify $\buvec{m}$ in spherical coordinates using polar angle $\gamma$ and azimuthal angle $\eta$, $\buvec{m} = (\cos\eta\sin\gamma, \sin\eta\sin\gamma, \cos\gamma)$. Thus the measurement Hamiltonian is 
\begin{align}\label{eq:1dvhjbbdhvbdhjv}
\op{H}_m(\bvec{x}) &= -   \mu \bopvecgr{\sigma} \cdot \bvec{B}_m(\bvec{x})
= -   \mu \frac{\beta q}{T} \bigl[ \cos\eta\sin\gamma \,\op{\sigma}_x \notag \\ & \quad\, + \sin\eta\sin\gamma \,\op{\sigma}_y + \cos\gamma \,\op{\sigma}_z\bigr].
\end{align}
The condition of a weak measurement corresponds to $T \gg \omega_0^{-1}$. If we think of $q$ as the one-dimensional position operator for the $\buvec{m}$ axis, we can see that this Hamiltonian generates changes in particle momentum along the $\buvec{m}$ direction \footnote{Since the operator $\op{q}$ does not commute with the Hamiltonian $\op{H}_p=\bopvec{p}^2/2m$ associated with the phase-space degree of freedom of the particle, it is not a constant of motion. Because this noncommutativity complicates the mathematical treatment without altering the possibility and physics of protective measurements \cite{Dass:1999:az}, we follow, without loss of generality, the common approach \cite{Aharonov:1993:jm,Dass:1999:az,Schlosshauer:2015:uu} of considering the particle in its rest frame, such that $\op{H}_p =0$.}. These momentum changes represent the pointer shifts in the model. 

Suppose that at $t=0$ (the start of the measurement interaction), the system $S$ is in the initial state $\ket{\psi(0)} = \ket{0}\ket{\Phi(p_0)}$, where $\ket{\Phi(p_0)} $ is the initial wave function for particle momentum along $\buvec{m}$, which we take to be a Gaussian of width $\sigma_p$ centered at $p_0$,
\begin{equation}\label{eq:18}
\Phi_{p_0}(p) = \braket{p}{\Phi(p_0)} = \left( \frac{1}{2\pi\sigma_p^2}\right)^{1/4} \exp \left[-\frac{(p-p_0)^2}{4\sigma_p^2}\right].
\end{equation}
In the weak-measurement limit $T \gg \omega_0^{-1}$ [i.e., $\abs{\bvec{B}_m(\bvec{x}) } \ll \abs{\bvec{B}_0}$], the state of $S$ at the conclusion of the measurement ($t=T$) is \cite{Aharonov:1993:jm,Schlosshauer:2015:uu}
\begin{align}\label{eq:29}
\ket{\psi(\bvec{x}, T)} &\approx \exp\left( \frac{\I \omega_0 T}{2} \right) \ket{0} \exp\left(\frac{\I \mu \beta q  \cos\gamma }{\hbar} \right) \ket{\Phi(p_0)}  \notag\\&= \exp\left( \frac{\I \omega_0 T}{2} \right) \ket{0} \ket{\Phi(p_0+\mu\beta\cos\gamma)}  .
\end{align}
Since $\cos\gamma = \bra{0} \bopvecgr{\sigma} \cdot \buvec{m}  \ket{0}$, this shows that the center of the momentum wave packet has shifted by an amount proportional to the expectation value of $\bopvecgr{\sigma} \cdot \buvec{m}$ in the initial spin state, while the spin state itself is left approximately undisturbed. By measuring this momentum change along $\buvec{m}$, the expectation value $\bra{0} \bopvecgr{\sigma} \cdot \buvec{m}  \ket{0}$ can be determined. This momentum change can be obtained by measuring the final position of the particle when it has completed its travel through the measurement field, giving the deflection of the particle in the direction $\buvec{m}$ \cite{Aharonov:1993:jm}. 

\subsection{\label{sec:envir-inter}Environmental interaction}

To study the influence of an environment and decoherence, we include the interaction of the spin degree of freedom of $S$ with an environment $E$ consisting of $N$ spin-$\frac{1}{2}$ particles. We take the system to couple to the environment through its $\op{\sigma}_x$ coordinate,
\begin{equation}\label{eq:3}
\op{H}_{SE}=\frac{1}{2} \op{\sigma}_x \otimes  \sum_{i=1}^N g_i  \op{\sigma}_x^{(i)} \equiv \frac{1}{2} \op{\sigma}_x \otimes \op{E}, 
\end{equation}
where the $g_i$ are coupling coefficients, and we neglect the internal dynamics of the environment ($\op{H}_E=0$). This type of environmental interaction was used in one of the first models of decoherence \cite{Zurek:1982:tv}. It has since been studied repeatedly and its relevance to a large class of physical situations has been emphasized \cite{Schliemann:2002:yy,Dobrovitski:2003:az,Cucchietti:2005:om,Schlosshauer:2005:bb}. 

An orthonormal set of eigenstates $\ket{E_n}$ (where $n=0,1,\hdots,2^N-1$) of the environment operator $\op{E}$ defined in Eq.~\eqref{eq:3} is given by tensor products $\ket{k_1}_x\ket{k_2}_x \cdots \ket{k_N}_x$, $k_i \in \{0,1\}$, of eigenstates of the individual environment spin operators $\op{\sigma}^{(i)}_x$, with eigenvalues 
\begin{equation}
\label{eq:1fssvgihTG8}
  \epsilon_n = \sum_{i=1}^N (-1)^{k_i} g_i.
\end{equation}
At $t=0$, we take the system--environment state to be in a pure product state,
\begin{equation}
\ket{\Psi(\bvec{x}, 0)} = \ket{0}\ket{\Phi(p_0)} \sum_{n=0}^{2^N-1} c_n\ket{E_n}.
\end{equation}
At $t=T$, the evolved state is
\begin{multline}\label{eq:9}
\ket{\Psi(\bvec{x}, T)} =\\=
\sum_{n=0}^{2^N-1} c_n \exp \left[ -\frac{\I}{\hbar} \left(\op{H}_S + \op{H}_m(\bvec{x}) + \frac{1}{2} \epsilon_n\op{\sigma}_x \right)T \right] \\ \times\ket{0}\ket{\Phi(p_0)} \ket{E_n} .
\end{multline}
Thus, for each environmental state $\ket{E_n}$ we can consider an effective system Hamiltonian \cite{Cucchietti:2005:om}
\begin{equation}\label{eq:22}
\op{H}_S^{(n)} = -\mu B_0 \op{\sigma}_z + \frac{1}{2}\epsilon_n \op{\sigma}_x \equiv -\mu B_0 \op{\sigma}_z -\mu b_n \op{\sigma}_x,
\end{equation}
where $\bvec{b}_n = b_n \buvec{x} = -\frac{\epsilon_n}{2\mu}\buvec{x}$ is the magnetic field associated with $\ket{E_n}$. We shall refer to the $\bvec{b}_n$ as the environment fields.

For a given $\ket{E_n}$, the total Hamiltonian [including the measurement field~\eqref{eq:measfield}] may then be written as 
\begin{equation}\label{eq:8}
\op{H}^{(n)}(\bvec{x})  = -  \mu \bopvecgr{\sigma} \cdot \bvec{B}^{(n)}(\bvec{x}),
\end{equation}
where  $\bvec{B}^{(n)}(\bvec{x})$ is the effective field felt by the spin particle, with components
\begin{subequations}\label{eq:26}
\begin{align}\label{eq:1}
B_x^{(n)}(\bvec{x}) &= \frac{\beta q}{T} \cos\eta\sin\gamma+b_n, \\
B_y(\bvec{x}) &= \frac{\beta q}{T} \sin\eta\sin\gamma, \\
B_z(\bvec{x}) &= B_0+\frac{\beta q}{T} \cos\gamma.
\end{align}
\end{subequations}
We define dimensionless field parameters $\xi(\bvec{x})=\frac{\beta q}{B_0 T}$ and $\tilde{b}_n=\frac{b_n}{B_0}$ that quantify the strength of the measurement and environment fields relative to the protection field strength $B_0$. Then we can write the magnitude of $\bvec{B}^{(n)}(\bvec{x})$ as $B^{(n)}(\bvec{x})= B_0 \chi_n(\bvec{x})$ with
\begin{align}\label{eq:4}
\chi_n(\bvec{x})& = \bigl[ 1 + \tilde{b}_n^2 + \xi (\bvec{x})^2 + 2  \tilde{b}_n \xi (\bvec{x}) \cos\eta\sin\gamma
\notag\\ &\quad + 2 \xi (\bvec{x}) \cos\gamma \bigr]^{1/2}.
\end{align}
The components of the unit vector $\buvec{r}_n(\bvec{x})$ specifying the direction of $\bvec{B}^{(n)}(\bvec{x})$ are given by
\begin{subequations}\label{eq:16hjkhjk}
\begin{align}
r_n^x(\bvec{x}) &= \frac{\xi (\bvec{x})\cos\eta\sin\gamma +   \tilde{b}_n}{\chi_n (\bvec{x})}, \label{eq:24}\\
r_n^y(\bvec{x}) &= \frac{\xi (\bvec{x})\sin\eta\sin\gamma}{\chi_n (\bvec{x})}, \label{eq:25}\\
r_n^z(\bvec{x}) &= \frac{1 + \xi (\bvec{x}) \cos\gamma}{\chi_n (\bvec{x})}.\label{eq:16}
\end{align}
\end{subequations}
Note that $r_n^z (\bvec{x})=\cos\theta_n (\bvec{x})$, where $\theta_n (\bvec{x})$ is the polar angle of $\bvec{B}^{(n)}(\bvec{x})$.

\subsection{Time evolution}

The eigenstates of the Hamiltonian $\op{H}^{(n)}(\bvec{x})$ [see Eq.~\eqref{eq:8}] are 
\begin{align}
\ket{\buvec{r}_n^+ (\bvec{x})} &= \cos\frac{\theta_n (\bvec{x})}{2}\ket{0} + \sin\frac{\theta_n (\bvec{x})}{2}\E^{\I \phi_n (\bvec{x})}\ket{1},\\
\ket{\buvec{r}_n^-(\bvec{x})} &= \sin\frac{\theta_n (\bvec{x})}{2}\ket{0} - \cos\frac{\theta_n (\bvec{x})}{2}\E^{\I \phi_n (\bvec{x})}\ket{1},
\end{align}
where $\theta_n (\bvec{x})$ and $\phi_n (\bvec{x})$ are the polar and azimuthal angles of the net field direction $\buvec{r}_n(\bvec{x})$ given by \eqref{eq:16hjkhjk}. Then the state~\eqref{eq:9} at $t=T$ can be evaluated to 
\begin{align}\label{eq:5}
\ket{\Psi(\bvec{x},T)} &=
\sum_{n=0}^{2^N-1} c_n \bigg[\cos\frac{\theta_n (\bvec{x})}{2}\E^{ \I \mu T B_0 \chi_n (\bvec{x})/\hbar }\ket{\buvec{r}_n^+ (\bvec{x})}  \notag \\ &\quad + \sin\frac{\theta_n (\bvec{x})}{2}\E^{ -\I \mu T B_0 \chi_n (\bvec{x})/\hbar }\ket{\buvec{r}_n^- (\bvec{x})}\bigg]\notag\\ &\quad \times \ket{\Phi(p_0)}\ket{E_n}.
\end{align}
In the following, we will omit the argument $\bvec{x}$ and associate the position coordinate $q$ with the measured location of the particle along $\buvec{m}$ at $t=T$ \cite{Aharonov:1993:jm}. Since the measurement is weak, we have $\xi \ll 1$ and can therefore expand $\chi_n$ [see Eq.~\eqref{eq:4}] to first order in $\xi$,
\begin{equation}
\chi_n \approx \sqrt{1 + \tilde{b}_n^2} + \xi \left(\frac{\cos\gamma + \tilde{b}_n \cos\eta\sin\gamma}{\sqrt{1 + \tilde{b}_n^2}}\right).
\end{equation}
Using this approximation from here on, the exponentials in Eq.~\eqref{eq:5} can be written as
\begin{equation}\label{eq:11}
\exp\left( \pm \frac{\I}{\hbar} \mu T B_0 \chi_n \right) = \exp\left( \pm \I \Omega_n T \right) \exp\left( \pm \frac{\I  q\Delta p_n}{\hbar} \right),
\end{equation}
where 
\begin{equation}
\Omega_n = \frac{\mu \sqrt{B_0^2 + b_n^2} }{\hbar},
\end{equation}
and
\begin{equation}\label{eq:10}
\Delta p_n = \mu\beta \left(\frac{\cos\gamma + \tilde{b}_n \cos\eta\sin\gamma}{\sqrt{1 + \tilde{b}_n^2}}\right)
\end{equation}
is the magnitude of the momentum change (pointer shift) in the direction $\buvec{m}$ of the measurement field. 

We see from Eq.~\eqref{eq:10} that the influence of the environment on the pointer shift amounts to additional momentum kicks. The equation shows that the influence is maximized for $\gamma=\frac{\pi}{2}$ and $\eta=0$, when the measurement field is oriented along the $x$ axis and therefore coincides with the orientation of the environment field. Note also that the environment influences the pointer shift even though it does not directly couple to the pointer variable itself but rather to the spin coordinate $\op{\sigma}_x$ [compare Eq.~\eqref{eq:3}]. One way to understand this behavior is by recalling that the environment, for each state $\ket{E_n}$, gives rise to an effective environment-modified Hamiltonian $\op{H}_S^{(n)}$ for the spin degree of freedom of the particle [see Eq.~\eqref{eq:22}].  From perturbation theory it follows that the effect of the measurement interaction on the pointer, treated as a small perturbation, is given (to first order) by the expectation value of the spin part  $\bopvecgr{\sigma} \cdot \buvec{m}$ of the perturbation in the eigenbasis of the unperturbed Hamiltonian $\op{H}_0$. Without an environment,  $\op{H}_0$ is equal to $\op{H}_S$ [Eq.~\eqref{eq:vshvbjfdjhvs}] with eigenbasis $\{\ket{0},\ket{1}\}$, and the expectation value of the perturbation in this basis is proportional to $\bra{0} \bopvecgr{\sigma} \cdot \buvec{m}  \ket{0}= \cos\gamma$, which is the familiar result~\eqref{eq:29}. In the presence of an environment, however, $\op{H}_0$ is represented by the family of Hamiltonians $\op{H}_S^{(n)}$. It follows that, for each state $\ket{E_n}$, the expectation value of the perturbation must now be evaluated in the eigenbasis of $\op{H}_S^{(n)}$, which yields the term in parentheses in Eq.~\eqref{eq:10}. Thus, the influence of the environment on the pointer shift may be understood as a consequence of the modification of the particle's spin Hamiltonian by the environment. 

Using Eq.~\eqref{eq:11}, the evolved state~\eqref{eq:5} can be written as $\ket{\Psi} = \sum_{n=0}^{2^N-1} c_n \ket{\psi^{(n)}}\ket{E_n}$ with
\begin{align}\label{eq:12}
\ket{\psi^{(n)}}&=\cos\frac{\theta_n }{2}\E^{ \I \Omega_n T }\ket{\buvec{r}_n^+}\ket{\Phi(p_0+\Delta p_n)}  \notag\\&\quad+ \sin\frac{\theta_n }{2}\E^{ -\I \Omega_n T}\ket{\buvec{r}_n^-}\ket{\Phi(p_0-\Delta p_n)}.
\end{align}
Here we have omitted the time argument $T$ in the state vector symbols for notational simplicity, as evaluation at $t=T$ will be implicit from here on. The corresponding reduced density matrix of the system $S$ is
\begin{equation}\label{eq:2}
\op{\rho}_{S} = \text{Tr}_E \, \ketbra{\Psi} {\Psi} =
\sum_{n=0}^{2^N-1} \abs{c_n}^2 \ketbra{\psi^{(n)}}{\psi^{(n)}},
\end{equation}
which is an incoherent mixture of the states~\eqref{eq:12}.

\section{\label{sec:infl-decoh-init}Influence of the environment on the spin state}

We first study the decoherence imparted by the environmental interaction on the spin state of the system. By tracing over the momentum degree of freedom in the density matrix~\eqref{eq:2}, we obtain the reduced density matrix $\op{\rho}$ for the spin degree of freedom at $t=T$,
\begin{align}
\op{\rho} &= \sum_{n=0}^{2^N-1} \abs{c_n}^2 \cos^2\frac{\theta_n}{2}\ketbra{\buvec{r}_n^+}{\buvec{r}_n^+}
+ \sin^2\frac{\theta_n}{2}\ketbra{\buvec{r}_n^-}{\buvec{r}_n^-}
\notag \\ &\quad + \Gamma_n \cos\frac{\theta_n}{2}\sin\frac{\theta_n}{2}\E^{2\I \Omega_n T} \ketbra{\buvec{r}_n^+}{\buvec{r}_n^-}
\notag \\ &\quad + \Gamma_n \cos\frac{\theta_n}{2}\sin\frac{\theta_n}{2}\E^{-2\I \Omega_n T} \ketbra{\buvec{r}_n^-}{\buvec{r}_n^+},\label{eq:23}
\end{align}
where $\Gamma_n = \braket{\Phi(p_0+\Delta p_n)} {\Phi(p_0-\Delta p_n)}$ measures the overlap of the momentum-shifted wave packets. To quantify the amount of disturbance of the initial spin state $\ket{0}$ caused by the measurement and environment fields, we calculate the probability $\mathcal{P}_1 = \bra{1}\rho\ket{1}$ of finding the system in the orthogonal state $\ket{1}$ at $t=T$. From Eq.~\eqref{eq:23}, this probability is
\begin{equation}\label{eq:17}
\mathcal{P}_1 = \frac{1}{2}\sum_{n=0}^{2^N-1} \abs{c_n}^2 \left\{ \sin^2\theta_n\left[1-\Gamma_n\cos(2 \Omega_n T) \right] \right\},
\end{equation}
where, from Eqs.~\eqref{eq:4} and \eqref{eq:16hjkhjk},
\begin{equation}\label{eq:27}
\sin^2\theta_n = \frac{\xi^2 \sin^2\gamma +   \tilde{b}^2_n + 2\tilde{b}_n \xi \cos\eta\sin\gamma}{1 + \tilde{b}_n^2 + \xi^2 + 2  \tilde{b}_n \xi \cos\eta\sin\gamma
 + 2 \xi \cos\gamma},
\end{equation}
and we have used that $\sum_{n=0}^{2^N-1} \abs{c_n}^2=1$. 

Note that even in the absence of decoherence, $\mathcal{P}_1$ is nonzero due to the presence of the measurement field \cite{Schlosshauer:2015:uu}. This can be seen from Eq.~\eqref{eq:17}, which, without environmental interactions, simplifies to 
\begin{equation}\label{eq:17dsc}
\mathcal{P}_1 = \frac{1}{2}\sin^2\theta\left[1-\Gamma \cos(\omega_0 T) \right].
\end{equation}
This probability oscillates as a function of $T$. However, because in a protective measurement the magnitude (i.e., $\omega_0$) of the protection field need not be known \cite{Aharonov:1993:jm,Dass:1999:az}, one has in general not enough information to choose $T$ such that $\mathcal{P}_1=0$ \cite{Schlosshauer:2015:uu}. Instead, we use Eq.~\eqref{eq:17dsc} to obtain an upper bound on $\mathcal{P}_1$, by replacing $\cos(\omega_0 T)$ by its minimum value $-1$ and also setting $\Gamma=1$, as both of these choices maximize $\mathcal{P}_1$. Then 
\begin{equation}\label{eq:17djkd44sc}
\mathcal{P}_1\le \sin^2\theta = \frac{\xi^2\sin^2\gamma}{1 + \xi^2 + 2 \xi \cos\gamma},
\end{equation}
where we have used Eq.~\eqref{eq:27} with $\tilde{b}_n=0$. $\mathcal{P}_1$ is largest for $\gamma=\frac{\pi}{2}$ when the protection and measurement fields are orthogonal. In this case, $\xi = 0.1$ gives  $\mathcal{P}_1 \le 0.01$, i.e., the probability of state disturbance due to the measurement field alone (without decoherence) is no greater than 0.01 for all possible orientations of the measurement field. From here on, we will use this value of $\xi$ as a reasonable choice for the strength of the measurement interaction.

We now return to the consideration of added decoherence as given by Eq.~\eqref{eq:17}, and rewrite this equation in equivalent integral form as
\begin{align}\label{eq:14}
\mathcal{P}_1 &= \frac{1}{2}\int_{-\infty}^\infty \D  \tilde{b} \, w(\tilde{b})  \left\{ \sin^2\theta(\tilde{b})\left[1-\Gamma(\tilde{b})\cos[2 \Omega(\tilde{b}) T]\right] \right\},
\end{align}
where $w(\tilde{b}) = \sum_{n=0}^{2^N-1} \abs{c_n}^2\delta(\tilde{b}-\tilde{b}_n)$ is the spectral density describing the distribution of the $\tilde{b}$. It has been shown \cite{Cucchietti:2005:om} that already for modest values of $N$ and for a large class of distributions of the couplings $g_i$ [Eq.~\eqref{eq:3}], the distribution of the energies $\epsilon_n$ given by Eq.~\eqref{eq:1fssvgihTG8}, and therefore also the distribution of the environment fields $\tilde{b}_n$, is well described by a Gaussian,
\begin{equation}\label{eq:15}
w(\tilde{b}) = \frac{1}{\sqrt{2\pi s_d^2}} \exp \left(-\frac{\tilde{b}^2}{2s_d^2}\right),
\end{equation}
where $s_d$ represents a typical strength of the environment field relative to the protection field strength $B_0$. We will use this distribution from here on. Also, in the regime $T \gg \Omega$ relevant to a protective measurement, Eq.~\eqref{eq:14} simplifies to
\begin{equation}\label{eq:13}
\mathcal{P}_1 \approx  \frac{1}{2} \int_{-\infty}^\infty \D  \tilde{b} \, w(\tilde{b})  \sin^2\theta(\tilde{b}),
\end{equation}
which establishes the first main result of this paper. 

\begin{figure}
\includegraphics[width=3.4in]{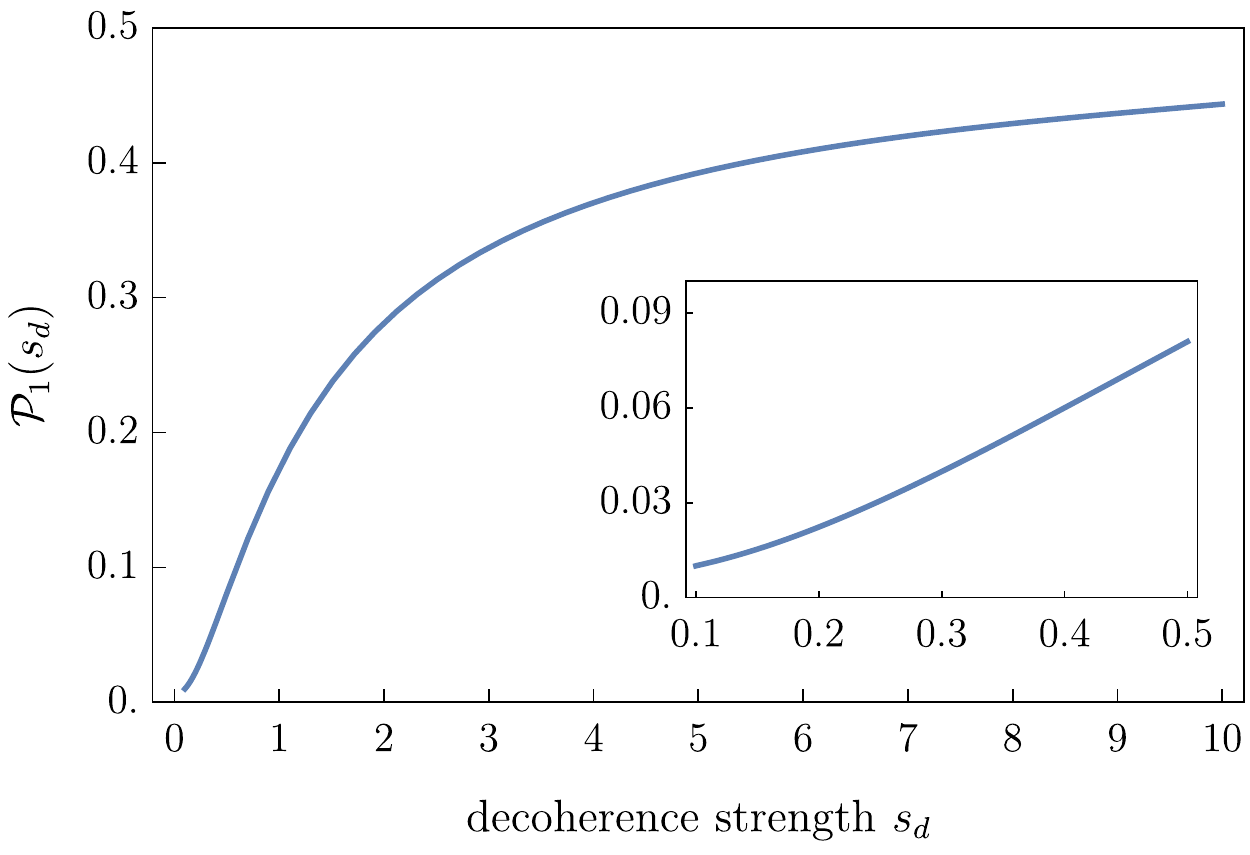} 
\caption{\label{fig:statedist}Influence of decoherence on the initial state. The plot gives the probability $\mathcal{P}_1$ [see Eq.~\eqref{eq:13}] of finding the system in the orthogonal spin state $\ket{1}$ at the conclusion of the protective measurement, shown as a function of the decoherence strength $s_d$. The inset magnifies the plot in the region of small values of $s_d$. The choices for the parameters of the measurement field are: strength  $\xi = 0.1$ and orientation $\gamma = \frac{\pi}{2}, \eta = 0$ (this orientation maximizes the influence of the environment fields).}
\end{figure} 

The probability $\mathcal{P}_1$ given by Eqs.~\eqref{eq:15} and \eqref{eq:13} is shown in Fig.~\ref{fig:statedist} as a function of the decoherence strength $s_d$. If the decoherence is very weak ($s_d \ll 1$), then for a typical value $\tilde{b} \approx s_d$ the net field will be close to the $z$ direction, i.e., $\sin^2\theta(s_d)\ll 1$. Thus Eq.~\eqref{eq:13} gives $\mathcal{P}_1 \ll 1$, showing that the initial state is not substantially affected by the presence of the environment. In the opposite limit where the environment fields are so strong as to dominate the evolution ($s_d \gg 1$), a typical net field will be close to the $x$ direction, i.e., $\sin^2\theta(s_d) \approx 1$, which yields $\mathcal{P}_1 \approx \frac{1}{2}$ from Eq.~\eqref{eq:13}. In this case, the environmental monitoring of the $\op{\sigma}_x$ spin coordinate leads to a loss of most of the coherence between the components $\ket{0}_x$ and $\ket{1}_x$ in the initial state $\ket{0}=\frac{1}{\sqrt{2}}(\ket{0}_x+\ket{1}_x)$, giving an approximately maximally mixed state and thus roughly equal probabilities of finding either $\ket{0}$ or $\ket{1}$. 

We can use this information to define two decoherence regimes. (i) We refer to \emph{weak decoherence} as the regime in which the presence of the environment does not contribute an appreciable probability of leaving the initial state. We choose $\mathcal{P}_1 \le 0.05$ as the upper limit for state disturbance in this regime, which corresponds to $s_d \le 0.35$. (ii) We refer to \emph{strong decoherence} as the regime $s_d \gtrsim 1$ where a typical strength of the environment field is on the order of (or exceeding) the size of the protection field, and significant state disturbance results (for $s_d = 1.0$ we have $\mathcal{P}_1 = 0.17$). Since the goal of a protective measurement is to leave the initial state approximately unchanged, only the regime of weak decoherence can be said to allow for a proper protective measurement. 

\section{\label{sec:infl-decoh-point}Influence of the environment on the pointer shift}

A second important consideration with respect to the quality of the protective measurement is the pointer shift. Therefore, we now turn to the question of how the pointer shift is influenced by the presence of the environment. By tracing over the spin degree of freedom of the system in the density matrix~\eqref{eq:2}, we obtain the reduced density matrix $\rho(p)$ for the momentum degree of freedom at $t=T$, 
\begin{align}\label{eq:19}
\rho(p) &= \int_{-\infty}^\infty  \D \tilde{b}\, w(\tilde{b}) \left[ \cos^2\frac{\theta(\tilde{b})}{2} \abs{\Phi_{p_0+\Delta p(\tilde{b})}(p)}^2 \right.\notag\\&\quad \left.+ \sin^2\frac{\theta(\tilde{b})}{2}  \abs{\Phi_{p_0-\Delta p(\tilde{b})}(p)}^2\right],
\end{align}
where we have again gone to the continuum limit using the Gaussian distribution $w(\tilde{b})$ given by Eq.~\eqref{eq:15}. This is an incoherent mixture of the Gaussian pointer states $\Phi_{p_0}(p)$ [see Eq.~\eqref{eq:10}] shifted in momentum by $\pm \Delta p(\tilde{b})$ as given by Eq.~\eqref{eq:10}. Explicitly, 
\begin{multline}\label{eq:20}
\abs{\Phi_{\pm\Delta p(\tilde{b})}(p)}^2 \equiv \abs{\Phi_\pm(p,\tilde{b})}^2 = \frac{1}{\sqrt{2\pi s_p^2}} \\ \quad\times\exp \left\{-\frac{1}{2\sigma_p^2} \left[ p \mp \mu\beta \left(\frac{\cos\gamma + \tilde{b} \cos\eta\sin\gamma}{\sqrt{1 + \tilde{b}^2}}\right) \right]^2\right\},
\end{multline}
where we have set $p_0=0$ for simplicity (we are concerned only with changes in momentum, and any nonzero $p_0$ merely adds a constant to the argument of the Gaussian).

Only the pointer shift $+\Delta p(\tilde{b})$, which corresponds to the first term in Eq.~\eqref{eq:19}, represents the correct shift that encodes the desired expectation value $\bra{0} \bopvecgr{\sigma} \cdot \buvec{m}  \ket{0}$, while the reversed shift $-\Delta p(\tilde{b})$ encodes the expectation value of $\bopvecgr{\sigma} \cdot \buvec{m}$ in the spin state $\ket{1}$ orthogonal to the initial state $\ket{0}$. However, in the case of weak decoherence ($s_d \ll 1$) relevant to protective measurements (see Sec.~\ref{sec:infl-decoh-init}), the first term in Eq.~\eqref{eq:19} dominates. This is so because to first order in $\xi$ and $\tilde{b}$, $\cos^2\frac{\theta(\tilde{b})}{2} = 1-\frac{1}{2}\xi \tilde{b}\cos\eta\sin\gamma$, and since the term $\frac{1}{2}\xi \tilde{b} \cos\eta\sin\gamma$ is of second order, we can neglect it. Then Eq.~\eqref{eq:19} becomes
\begin{equation}\label{eq:1lfdk9}
\rho(p) = \int_{-\infty}^\infty  \D \tilde{b}\, w(\tilde{b}) \abs{\Phi_+(p,\tilde{b})}^2.
\end{equation}
Still working with the case of weak decoherence, we expand the pointer shift in the argument of the exponential~\eqref{eq:20} to first order in $\tilde{b}$,
\begin{multline}\label{eq:21}
\abs{\Phi_\pm(p,\tilde{b})}^2 = \frac{1}{\sqrt{2\pi \sigma_p^2}}\\\times\exp \left\{-\frac{1}{2\sigma_p^2} \left[ p \mp \mu\beta \left(\cos\gamma + \tilde{b} \cos\eta\sin\gamma\right) \right]^2\right\}.
\end{multline}
We will now evaluate the state $\rho(p)$ given by Eqs.~\eqref{eq:1lfdk9} and \eqref{eq:21}. Introducing the dimensionless momentum variable $\tilde{p}=p/\mu\beta$ and defining $\tilde{b}'=\tilde{b} \cos\eta\sin\gamma$, we rewrite  Eq.~\eqref{eq:1lfdk9} in the form
\begin{equation}\label{eq:1lfdjhvfjkhk9}
\rho(\tilde{p}) = \int_{-\infty}^\infty  \D \tilde{b}'\, u(\tilde{b}') v ( \tilde{p}-\tilde{b}').
\end{equation}
Here
\begin{equation}
u(\tilde{b}') = \frac{1}{\sqrt{2\pi s_u^2}} \exp \left(-\frac{\tilde{b}'^2}{2s_u^2}\right)
\end{equation}
is the Gaussian distribution~\eqref{eq:15} transformed to the variable $\tilde{b}'$, with mean $\mu_u = 0$ and width $\sigma_u=s_d \cos\eta\sin\gamma$, and [compare Eq.~\eqref{eq:21}]
\begin{equation}
v(\tilde{b}') = \frac{1}{\sqrt{2\pi \sigma_{\tilde{p}}^2}}\exp \left[-\frac{1}{2 \sigma_{\tilde{p}}^2} \left( \tilde{b}' -\cos\gamma \right)^2\right]
\end{equation}
is a Gaussian with mean $\mu_{\tilde{p}} = \cos\gamma$ and width $\sigma_{\tilde{p}}=\sigma_p/\mu\beta$. Therefore, Eq.~\eqref{eq:1lfdjhvfjkhk9} is a convolution of two Gaussians in the free variable $\tilde{p}=p/\mu\beta$, with mean
\begin{equation}\label{eq:63684tf3g}
\mu=\mu_{\tilde{p}}+\mu_u=\cos\gamma
\end{equation}
and variance 
\begin{equation}\label{eq:6}
\sigma^2 = \sigma_{\tilde{p}}^2+\sigma_u^2=(\sigma_p/\mu\beta)^2 + (s_d \cos\eta\sin\gamma)^2.
\end{equation}

Equations~\eqref{eq:63684tf3g} and \eqref{eq:6} establish the second main result of this paper.  They show that the center of the momentum probability distribution (with momentum expressed in the dimensionless variable $\tilde{p}=p/\mu\beta$) still shifts by $\cos\gamma$ just as without an environment present, but that the interaction with the environment broadens the distribution through the term $(s_d \cos\eta\sin\gamma)^2$. Note that the broadening depends both on the strength $s_d$ of the environmental interaction and on the orientation $(\gamma, \eta)$ of the measurement field. It diminishes the accuracy with which the expectation value $\bra{0} \bopvecgr{\sigma} \cdot \buvec{m}  \ket{0}$ can be inferred from a measurement of the particle's momentum change in the $\buvec{m}$ direction. Thus, the interaction with the environment leads to a smearing-out of the pointer and acts as noise on the pointer shift. 

\begin{figure}
\includegraphics[width=3.4in]{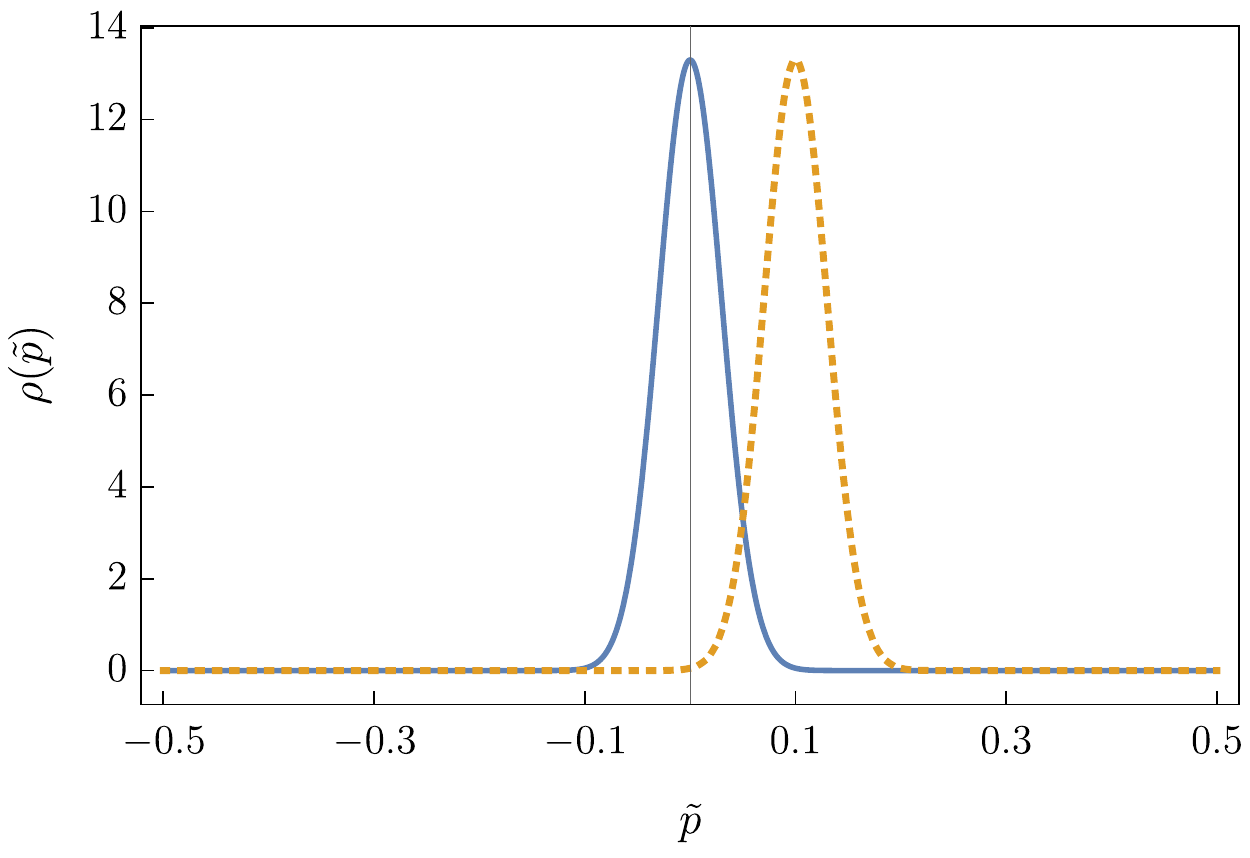} 
\caption{\label{fig:res}Evolution of the initial probability distribution for the pointer momentum (solid line) to the final momentum-shifted probability distribution (dotted line) in the absence of an environment, shown as a function of the dimensionless momentum variable $\tilde{p}=p/\mu\beta$. The width $\sigma_{\tilde{p}}$ of the momentum probability distribution was chosen to be $\sigma_{\tilde{p}}=0.03$ to enable adequate resolution of a shift of size 0.1 as shown.}
\end{figure}

We will now explore the effect of the broadening. First, we need to choose a reasonable value for the width $\sigma_{\tilde{p}}$ of the initial momentum wave packet (we will measure momentum in terms of $\tilde{p}=p/\mu\beta$ from here on). The size of the pointer shift $\cos\gamma = \bra{0} \bopvecgr{\sigma} \cdot \buvec{m}  \ket{0}$ varies between 0 and 1, so let us suppose that we would like to resolve pointer changes of size 0.1 (this corresponds to variations in $\gamma$ of up to $5^\circ$). As seen from Fig.~\ref{fig:res}, in the absence of environmental interactions the choice $\sigma_{\tilde{p}}=0.03$ offers a good distinction of the original Gaussian wave packet $\Phi(\tilde{p})$ from the momentum-shifted wave packet  $\Phi(\tilde{p}-0.1)$, giving an overlap of less than 0.1.

\begin{figure}
\includegraphics[width=3.4in]{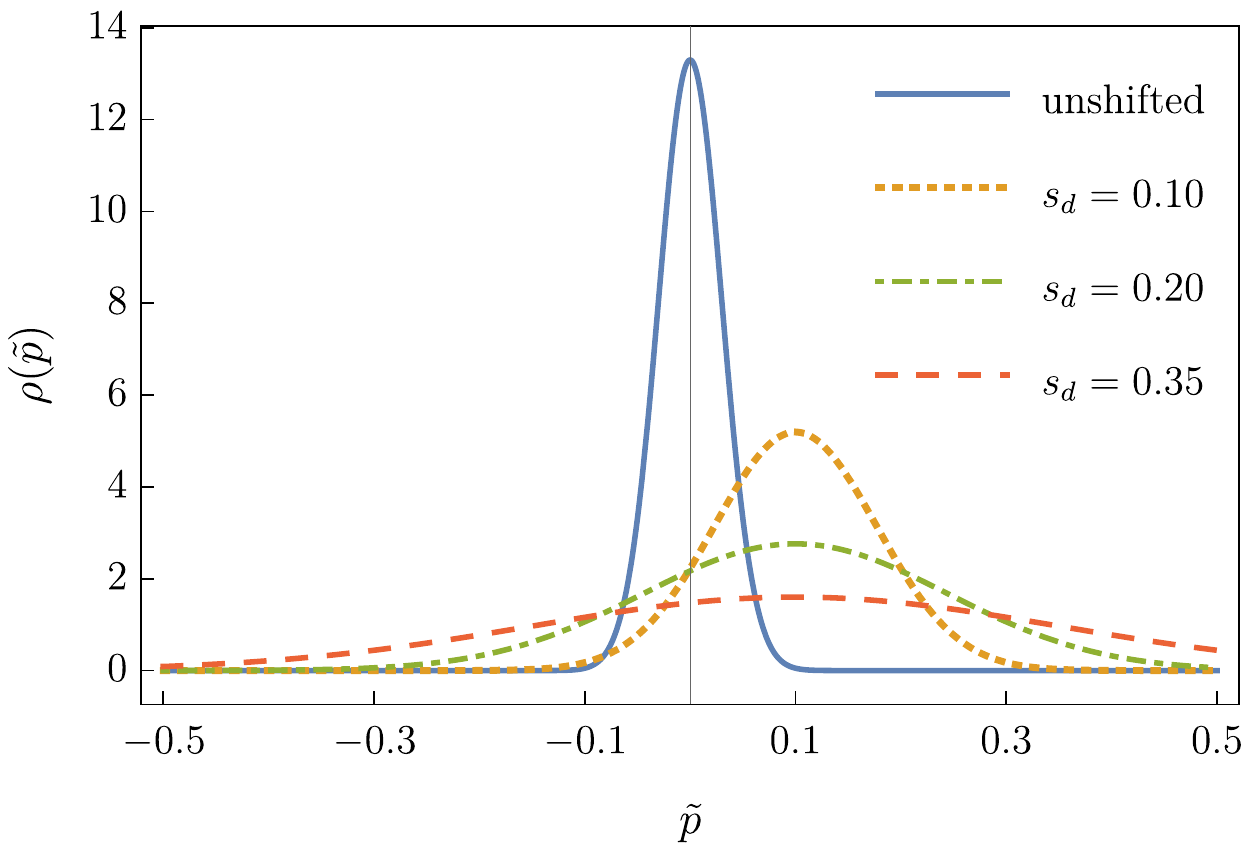} 
\caption{\label{fig:broad}Broadening of the final probability distribution $\rho(\tilde{p})$ for the pointer momentum due to the interaction with the spin environment, shown for different weak decoherence strengths $s_d$ and as a function of the dimensionless momentum variable $\tilde{p}=p/\mu\beta$. The initial (unshifted) momentum distribution is shown for reference. The chosen orientation of the measurement field is $\gamma=\frac{\pi}{4}$ and $\eta=0$.}
\end{figure}
 
Figure~\ref{fig:broad} shows the environment-induced broadening of the probability distribution [as given by Eq.~\eqref{eq:6}] for the pointer momentum at the conclusion of the measurement, for different decoherence strengths $s_d$ within the weak decoherence regime. We see that even for such weak decoherence when the initial spin state remains largely unaffected by the presence of the environment, a significant broadening of the pointer's momentum distribution occurs. For example, for $s_d=0.20$ (which corresponds to $\mathcal{P}_1 \approx 0.02$ and thus only insignificant disturbance of the spin state), the distribution has become so wide as to make all but impossible the reliable estimation of $\cos\gamma = \bra{0} \bopvecgr{\sigma} \cdot \buvec{m}  \ket{0}$ from the measured particle momentum in the $\buvec{m}$ direction. This indicates that the chief detrimental influence of the environment on a protective measurement arises in the form of washing-out of the pointer probability distribution associated with the pointer shift. It leads to a substantial reduction in the accuracy with which the desired expectation value can be measured and, as we have seen, is a significant factor even when the state of the system is not appreciably affected by the environment.

\section{\label{sec:envir-effects-with}Environmental effects without state disturbance}

As discussed in Sec.~\ref{sec:infl-decoh-init}, in the strong-decoherence regime, and with the relative orientation~\eqref{eq:22} of the protection and environment fields, the initial spin state of the system will be substantially perturbed. Therefore, one of the two conditions of a proper protective measurement, i.e., that the initial state remains essentially unchanged in the course of the measurement, is violated. On the other hand, the influence of decoherence on a given quantum state depends also on the choice of the state (with some states being immune to decoherence \cite{Zurek:1982:tv}). For example, if the environment fields act along the $z$ direction and thus along the axis of the protection field (i.e., if the system couples to the environment via the $\op{\sigma}_z$ coordinate), they cannot disturb the initial spin state $\ket{0}$ since now the system starts out in an eigenstate of the system--environment interaction Hamiltonian. This will be true even when the environment dominates the evolution ($s_d \gg 1$). However, in this limit no pointer shift will occur, as can be seen from the following argument. 

If the environment fields are along the $z$ axis, then the components of the net fields $\bvec{B}^{(n)}$ are as in Eq.~\eqref{eq:26} but with the $b_n$ term now associated with the $z$ component. The magnitude of $\bvec{B}^{(n)}$ is $B_0 \chi_n$ with $\chi_n= \left( 1 + \tilde{b}_n^2 + \xi^2 + 2  \tilde{b}_n + 2  \tilde{b}_n \xi \cos\gamma + 2 \xi \cos\gamma \right)^{1/2}$. Expanding $\chi_n$ to first order in $\xi$, we obtain
\begin{equation}
\chi_n \approx \abs{1 + \tilde{b}_n} + \xi \cos\gamma\frac{1 + \tilde{b}_n}{\abs{1 + \tilde{b}_n}},
\end{equation}
which gives a pointer shift $\pm \mu \beta \cos\gamma$, where the sign is negative if $\tilde{b}_n<-1$ (i.e., if $b_n<-B_0$). 

Therefore, there is no broadening of the probability distribution for the pointer momentum, but whenever $\tilde{b}_n<-1$ we get a reversed pointer shift  $- \mu \beta \cos\gamma = \mu \beta\bra{1} \bopvecgr{\sigma} \cdot \buvec{m} \ket{1}$. This behavior is readily understood by noting that each environment field $b_n$ can be thought of as an added value to the protection field. Whenever $B_0+b_n \ge 0$, only the strength of the protection field is modified, but since the size of the pointer shift does not depend on this strength, the environment field does not affect the pointer shift. Whenever $B_0+b_n < 0$, however, the environment field modifies not only the strength but also the direction of the protection field, as the sum of the two fields is now in the $-z$ direction. With respect to this new direction, the initial spin state $\ket{0}$ becomes the higher-energy (excited) state, which is equivalent to using the orthogonal state $\ket{1}$ for the original $+z$ direction of the unmodified protection field, and thus the pointer shift will be proportional to $\bra{1} \bopvecgr{\sigma} \cdot \buvec{m} \ket{1}$. 

\begin{figure}
\includegraphics[width=3.4in]{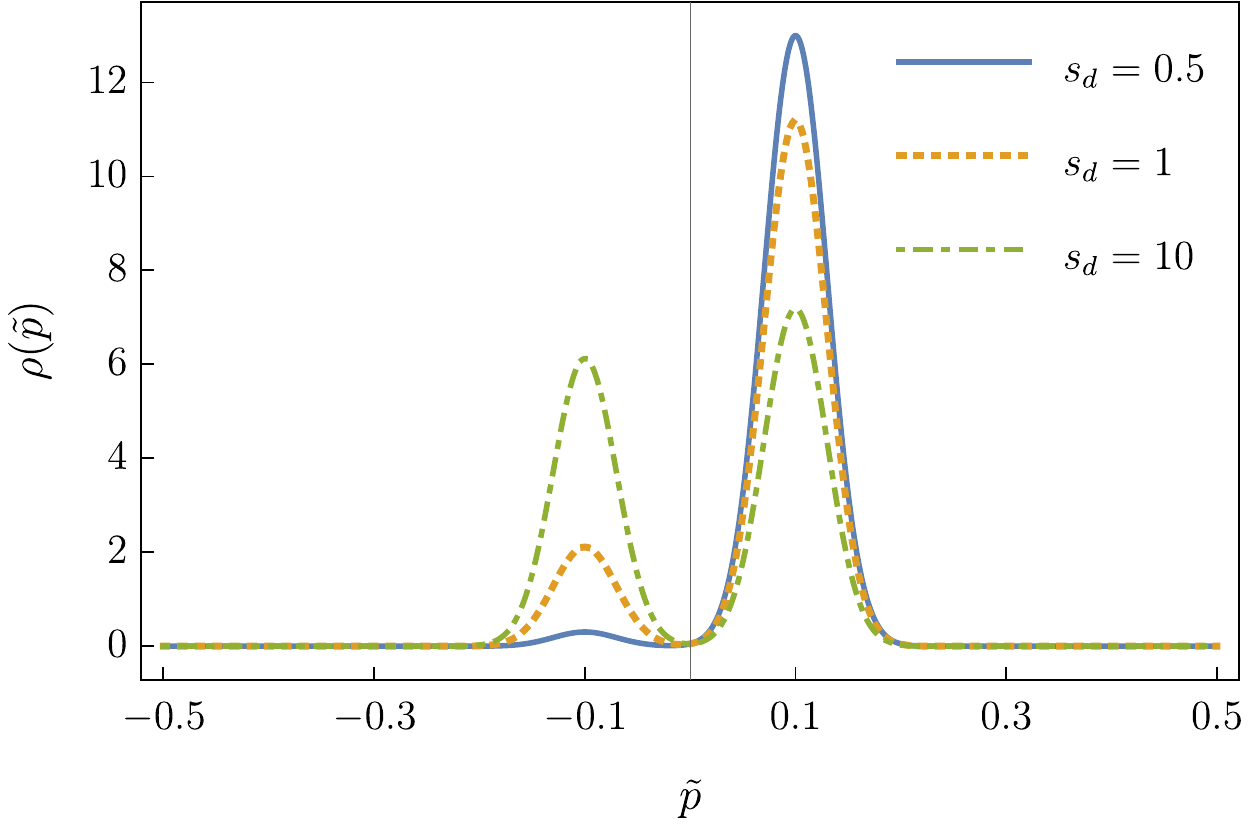} 
\caption{\label{fig:strong}Final probability distribution $\rho(\tilde{p})$ for the pointer momentum variable $\tilde{p}$ when the environment fields act along the axis of the protection field and thus do not disturb the initial spin state, shown for different decoherence strengths $s_d$ and as a function of the dimensionless momentum variable $\tilde{p}=p/\mu\beta$. The peaks in the region $\tilde{p}<0$ correspond to the reversed pointer shift $\Delta\tilde{p}=- \cos\gamma$.}
\end{figure}

Applying Eq.~\eqref{eq:19} to this situation (with $\theta \approx 0$, since the net field is close to the $z$ direction), the pointer state $\rho(p)$ can be written as
\begin{equation}
\rho(p) \approx \mathcal{P}_+\abs{\Phi_{p_0+\mu\beta\cos\gamma}(p)}^2 + (1-\mathcal{P}_+)\abs{\Phi_{p_0-\mu\beta\cos\gamma}(p)}^2,
\end{equation}
where $\mathcal{P}_+ = \int_{-1}^\infty  \D \tilde{b}\, w(\tilde{b})$ is the probability of getting $\tilde{b} > -1$ and hence of obtaining the correct pointer shift $+ \mu \beta \cos\gamma$. In the weak-decoherence limit $s_d \ll 1$, $\mathcal{P}_+$ will be very close to 1 and thus the protective measurement will realize as if no environment were present, i.e., the environment will impart neither a disturbance of the initial state nor a change to the evolution of the pointer wave packet. Conversely, in the strong-decoherence regime $s_d \gtrsim 1$, $\mathcal{P}_+$ is substantially smaller than 1 and will approach $\frac{1}{2}$ for $s_d \gg 1$. As illustrated in Fig.~\ref{fig:strong}, this means that there is now a sizable likelihood of measuring a momentum value that corresponds to the reversed shift $- \mu \beta \cos\gamma$. Thus, even though the environment does not disturb the state of the system, the amount of information pertaining to the desired expectation value $\bra{0}\bopvecgr{\sigma} \cdot \buvec{m}\ket{0}$ that can be extracted from the protective measurement decreases as the decoherence strength is increased. In the limit  $s_d \gg 1$, the expectation value of the pointer momentum will be zero (compare Fig.~\ref{fig:strong}) and thus the pointer will encode no information about $\bra{0}\bopvecgr{\sigma} \cdot \buvec{m}\ket{0}$.

\section{\label{sec:stern-gerl-exper}Experimental scheme}

We will now discuss a possible approach to exploring our model in an experiment of the Stern--Gerlach type. First, recall that our results show that the phenomenological influence of the environment on the motional state of the spin particle is to impart noise in the form of momentum kicks. This can be seen directly from the final pointer state given by Eqs.~\eqref{eq:1lfdk9} and \eqref{eq:21}. In this incoherent mixture of momentum-space wave packets, each packet in the mixture is momentum-shifted by the combination of the system expectation value $\bra{0} \bopvecgr{\sigma} \cdot \buvec{m}  \ket{0}=\cos\gamma$ and a contribution from a random field $b$, which represents a portion of the effect of the interaction with the spin environment in terms of a local magnetic field. The distribution of these wave packets is given by the distribution $w(b)$ of the fields $b$ [see Eq.~\eqref{eq:15}]. The distribution $w(b)$ has zero mean, which implies that the momentum kicks average to zero and leave the mean of the pointer momentum unchanged, but the finite width of the distribution means that, as we have seen, the momentum distribution of the pointer becomes significantly broadened. 

These results suggest the following experimental scheme. We add a magnetic field $b$ to the Stern--Gerlach setup for the protective measurement [as described by Eqs.~\eqref{eq:vshvbjfdjhvs} and \eqref{eq:measfield}], oriented along the $x$ direction and randomly chosen from the Gaussian distribution $w(b)$. After passage of the spin particle (the atom) through the field, we measure, as usual, the pointer momentum shift in the direction $\buvec{m}$ of the measurement field~\eqref{eq:measfield}. As mentioned, this can be done by measuring the total displacement of the spin particle in this direction when the particle has reached the end of the measurement region, with the atomic position measured, for example, by shining a weak-intensity laser beam on the atom \cite{Aharonov:1993:jm}. The momentum kick delivered by the field $b$ will influence the displacement, and by repeating the experiment many times, the distribution of final pointer momenta along $\buvec{m}$ can be reconstructed and compared to the measured distribution in the absence of the added fields. Effectively, this procedure generates the momentum density matrix~\eqref{eq:1lfdk9} in terms of a physical ensemble of different noisy realizations of the atomic evolution. 
 
We now give some numerical estimates for the typical strength of the added fields $b$ and the resulting change in displacement of the spin particle.  We first discuss the relevant parameter values in the absence of an environment \cite{Schlosshauer:2015:uu}. For a momentum shift $\Delta p = \mu\beta  \cos\gamma$, the corresponding force on the atom caused by the measurement field is $F = \mu(\beta /T) \cos\gamma$, where $\beta/T$ is the magnitude of the gradient $\boldnabla B_m=\frac{\beta}{T} \buvec{m}$ of the measurement field given by Eq.~\eqref{eq:measfield}. In a modern realization of a Stern--Gerlach setup based on evaporated potassium atoms ($\mu = \unit[9.3 \times 10^{-24}]{J/T}$) \cite{Daybell:1967:sg},  the atoms are emitted from an oven at a typical temperature of $T_\text{oven}=\unit[420]{K}$, which translates to a most probable velocity of $v=\sqrt{2 k_BT_\text{oven}/m} \approx \unit[420]{m/s}$. The inhomogeneity in the direction of $\boldnabla B_m$ causes a spatial displacement given by 
\begin{align}\label{eq:28}
\Delta s &= \frac{\mu\beta \cos\gamma}{2m T} T^2 = \frac{\mu \abs{\boldnabla B_m}\cos\gamma}{2m} \left(\frac{d}{v}\right)^2 \notag\\&= \frac{\mu \abs{\boldnabla B_m}\cos\gamma}{4 k_BT_\text{oven}} d^2,
\end{align}
where $d$ is the size of the region containing the inhomogeneous measurement field. For $d=\unit[0.1]{m}$, $\gamma =\pi/4$, and a measurement-field gradient of $\boldnabla B_m \approx \unit[40]{T/m}$ (a typical value in a Stern--Gerlach experiment), the spatial displacement in the direction of the inhomogeneity is $\Delta s_0 \approx \unit[0.11]{mm}$. Note that the field parameter $\xi=\frac{\beta q}{B_0 T}$ (see Sec.~\ref{sec:envir-inter}) is here given by $(\boldnabla B_m) d/B_0$. For the values just stated and a protection field $B_0$ on the order of $\unit[10]{T}$, we have $\xi=0.4$, which corresponds to an upper limit on the state disturbance (due to the measurement field only) of 7\% [see Eq.~\eqref{eq:17djkd44sc}]. This indicates that with these parameter choices, one is able to fulfill the condition that the protective measurement leave the state of the system largely unchanged. Since a spatially extended uniform magnetic field of such strength may be difficult to realize experimentally, one can alternatively use a smaller field if the size $d$ of the measurement region is correspondingly enlarged. For example, for $d=\unit[1]{m}$ and the same displacement as before, the required measurement-field gradient is $\boldnabla B_m \approx \unit[0.04]{T/m}$. Then obtaining the same low state disturbance as before requires a uniform field strength of $B_0=\unit[1]{T}$. 

We can now include the random magnetic fields $b$ that produce the effect of the spin environment. In Sec.~\ref{sec:infl-decoh-init} we showed that in the regime of weak decoherence relevant to protective measurement, a threshold of 5\% for the disturbance of the spin state translates into an upper limit of $s_d \le 0.35$. In Sec.~\ref{sec:infl-decoh-point} we found that values in the range of $0.1 \le s_d \le 0.2$ already produce a substantial broadening of the pointer. Let us choose $s_d=0.2$, together with the value $B_0=\unit[1]{T}$ for the uniform field as discussed in the previous paragraph. Then we can experimentally produce the environmental broadening of the pointer momentum distribution by adding to the measurement region, in each iteration of the experiment, a random magnetic field drawn from $w(b)$ with width $B_0s_d =\unit[0.2]{T}$ (which represents a typical field strength). For this strength $b=B_0s_d$, the force on the atom is now $F = \mu(\beta /T) (\cos\gamma + s_d \sin\gamma)$ [see Eq.~\eqref{eq:21} with $\eta=0$], and the corresponding displacement is $\Delta s_1 \approx \unit[0.14]{mm}$, a 24\% difference compared to the displacement in the absence of the field. 

Thus, if we perform repeated runs of the experiment and plot the distribution of the resulting displacements, the distribution can be expected to follow a Gaussian of width $\Delta s_1-\Delta s_0$ (we assume that the spread of initial momenta of the atomic beam, as well as any free spreading, is sufficiently small such that the change in displacement induced by the added fields can be resolved). By comparing this distribution to the distribution obtained without the added fields, the effect of the simulated environment can be experimentally verified. Additionally, by varying the average strength of the added fields as quantified by $s_d$, changes in the width of the distribution can be observed. In this way, the dependence of the pointer broadening on the strength of the environmental interaction can be measured. 

An experiment of this kind could also be implemented using cold atoms \cite{Dass:1999:le}. Numerical estimates of the relevant parameters given in Ref.~\cite{Dass:1999:le} suggest that, provided low atomic velocities on the order of \unit[1]{cm/s} can be achieved, a much weaker protection field ($B_0 \approx \unit[1]{G}$) and a measurement strength of around $\xi=0.1$ will suffice to produce a measurement-induced beam displacement well in excess of both initial and free momentum spreading. 

\section{\label{sec:gener-appl}General protective qubit measurements}

So far, we have couched our analysis in the context of a setting of the Stern--Gerlach type. However, as already briefly indicated in Sec.~\ref{sec:model}, the model and the resulting calculations we have presented in this paper are generic to any protective measurement of a qubit. To see this, consider the Hamiltonian~\eqref{eq:7} together with the environmental contribution~\eqref{eq:3}, 
\begin{align}\label{eq:t677}
\op{H} &= \op{H}_S+\op{H}_m+\op{H}_{SE}\notag\\&=\frac{1}{2}\hbar \omega_0 \op{\sigma}_z + \frac{\zeta}{T} (\bopvecgr{\sigma} \cdot \buvec{m}) \otimes \op{K}_A +\frac{1}{2} \op{\sigma}_x \otimes  \sum_{i=1}^N g_i  \op{\sigma}_x^{(i)},
\end{align}
where we have written $\kappa(t)=\zeta/T$ for $t \in [0,T]$, with $\zeta$ a constant. This is the general form of the Hamiltonian describing the dynamics of a protective measurement of an arbitrary observable $\op{O}_S=\bopvecgr{\sigma} \cdot \buvec{m}$ on a generic qubit system $S$, with the apparatus pointer represented by an arbitrary observable $\op{K}_A$ that generates the pointer shift, and with $S$ coupled to an environment $E$ of two-level systems. For each environmental state $\ket{E_n}$ as defined in Sec.~\ref{sec:model}, the Hamiltonian~\eqref{eq:t677} can be equivalently mapped onto the Hamiltonian for a spin-$\frac{1}{2}$ particle interacting with an effective magnetic field as in Eq.~\eqref{eq:8}, i.e., $\op{H}^{(n)}(k)  = \bopvecgr{\sigma} \cdot \bvec{B}^{(n)}(k)$, where $k$ is the variable associated with the pointer operator $\op{K}_A$. The components of $\bvec{B}^{(n)}(k)$ are as in Eq.~\eqref{eq:26} but with straightforward substitutions of variables to match the variables used in the Hamiltonian~\eqref{eq:t677},
\begin{subequations}
\begin{align}
B_x^{(n)}(k) &= \frac{\zeta k}{T} \cos\eta\sin\gamma+\frac{1}{2}\epsilon_n, \\
B_y(k) &= \frac{\zeta k}{T} \sin\eta\sin\gamma, \\
B_z(k) &= \frac{1}{2}\hbar \omega_0 +\frac{\zeta k}{T} \cos\gamma,
\end{align}
\end{subequations}
where $\epsilon_n$ is the eigenvalue associated with $\ket{E_n}$.

It follows that the calculations and results of Secs.~\ref{sec:model}--\ref{sec:envir-effects-with} directly carry over to the general scenario described by the Hamiltonian~\eqref{eq:t677}. All that is required is to express the relevant variables in terms of the quantities used in the Hamiltonian~\eqref{eq:t677}. The dimensionless field parameters $\xi$ and $\tilde{b}_n$ defined in Sec.~\ref{sec:envir-inter} are now given by $\xi = 2\zeta k (\hbar \omega_0 T)^{-1}$ and $\tilde{b}_n=\epsilon_n(\hbar\omega_0)^{-1}$. As before, $\xi$ represents the relative sizes of $\op{H}_m$ and $\op{H}_S$ (i.e., the measurement strength), $\tilde{b}_n$ represents the relative sizes of $\op{H}_{SE}$ and $\op{H}_S$ for a given $\ket{E_n}$, and the width $s_d$ of the Gaussian distribution $w(\tilde{b})$ [see Eq.~\eqref{eq:15}] represents a typical value of the strength of the environmental interaction relative to $\op{H}_S$. The pointer is prepared in a Gaussian wave packet in the variable $\ell$ conjugate to $k$, with width $\sigma_\ell$. We can then apply Eqs.~\eqref{eq:63684tf3g} and \eqref{eq:6} with the substitutions $p \rightarrow \ell$ and $\mu\beta \rightarrow \zeta$. Analogous to the Stern--Gerlach case, this establishes the result that the center of the pointer wave packet is shifted (in the variable $\ell$) by an amount $\zeta\cos\gamma=\zeta \bra{0} \bopvecgr{\sigma} \cdot \buvec{m}  \ket{0}$, while the environment broadens the initial pointer wave packet so that its final variance is
\begin{equation}\label{eq:30}
\sigma^2 = \sigma_\ell^2 + (\zeta s_d \cos\eta\sin\gamma)^2.
\end{equation}

As a concrete example, consider the typical measurement setting in which the pointer operator $\op{K}_A$ [see Eq.~\eqref{eq:t677}] is the momentum operator generating spatial translations of a physical apparatus pointer, with the pointer initially represented by a Gaussian wave packet in position space. Equation~\eqref{eq:30} then quantifies the broadening of the distribution of final pointer positions due to the environment. The broadening implies an increase in the uncertainty in the measurement of the position of the pointer, and therefore an increased uncertainty in the outcome of the protective measurement, i.e., in the expectation value of the measured qubit observable.

The physical representation of the apparatus pointer depends of course on the specific experimental setting. The coupling between a qubit and an apparatus is a ubiquitous task in the control and readout of qubit systems in quantum information processing \cite{Nielsen:2000:tt}, and accordingly a large number of experimental realizations of such interactions exist, including weak measurement and quantum nondemolition schemes for systems such as superconducting quantum circuits \cite{Wendin:2017:aa,Vijay:2012:oo,Qin:2017:kk,Reuther:2011:zz}, quantum dots \cite{Jordan:2007:uu}, and ion traps \cite{Bruzewicz:2019:aa,Pan:2019:uu}, all of which might be adaptable to a future implementation of a protective measurement. In many of these cases, the apparatus can be modeled as a quantum resonator (harmonic oscillator), and the interaction can be tuned to the weak-coupling regime \cite{Qin:2017:kk,Jordan:2007:uu,Vijay:2012:oo,Pan:2019:uu}. For example, in superconducting quantum circuits \cite{Wendin:2017:aa,Qin:2017:kk,Reuther:2011:zz} in which a superconducting qubit is coupled to a transmission line resonator, the pointer is represented by an appropriate resonator mode (voltage for charge qubits, current for flux qubits), and the Hamiltonian for the measurement interaction takes the form $\op{H}_m = g \op{\sigma}_z(\op{a}+\op{a}^\dagger)$. Thus, the pointer observable is given by $\op{X}=\op{a}+\op{a}^\dagger$, generating measurable shifts in the conjugate quantity \cite{Wendin:2017:aa,Reuther:2011:zz}.

\section{\label{sec:discussion}Discussion and conclusions}

For a quantum measurement to be considered protective, it must leave the initial state of the system approximately unchanged while transferring information about the expectation value of an arbitrary system observable to the apparatus pointer. Our results show that while, unsurprisingly, interactions with a decoherence-inducing environment during the measurement make it harder to fulfill the first condition of minimal state disturbance, it is really the second condition of a faithful pointer shift that is most dramatically affected by the presence of the environment. For even when the system couples only weakly to the environment and the initial state does not become appreciably decohered, the probability distribution of the position of the pointer may be broadened so substantially as to make it difficult to reliably infer information about the expectation value of interest from a measurement of the pointer position. In this way, the environment acts as a significant source of noise on the pointer. 

Moreover, we have shown that the environment can have an effect on the pointer even when it does not lead to any decoherence of the state of the system. Specifically, it increases the likelihood of reading from the final pointer measurement a value that gives the expectation value not in the initial state of the system as desired, but in a state orthogonal to it. This can dramatically affect the fidelity of any quantum-state reconstruction based on protective measurements \cite{Schlosshauer:2015:uu}. 

We have also described how our model could be experimentally explored with the help of a setup of the Stern--Gerlach type. Here the influence of the environment can be simulated in terms of repeated noisy realizations of a standard Stern--Gerlach-type protective measurement augmented by random magnetic fields. The resulting pointer state will be equivalent to that obtained from actual spin--spin interactions between the qubit and the environmental spins. Such interactions would be difficult to realize in a controlled manner for an atom traversing the Stern--Gerlach apparatus. By contrast, the scheme we have outlined can be readily implemented once the protective measurement itself is experimentally available. 

In general, protective measurements, owing to the long duration of the measurement interaction, will be much more susceptible to couplings to an environment than short impulsive or weak measurements. Of course, to what extent decoherence plays a role in a specific experimental implementation of a protective measurement will depend on whether and how the measured system interacts with its environment. For experiments of the Stern--Gerlach type or for experiments based on photon polarization \cite{Piacentini:2017:oo}, unwanted environmental interactions may be reasonably easily controlled and minimized. This is unlikely to be the case in other potentially relevant physical situations, such as trapped ions \cite{Bruzewicz:2019:aa} or superconducting qubits \cite{Wendin:2017:aa,Reuther:2011:zz}. Indeed, given their ability to implement carefully controlled interactions between a qubit and an apparatus, the various experimentally studied qubit architectures for quantum information processing constitute promising platforms for the realization of a protective measurement. Since environmental interactions play a significant role in these qubit systems \cite{Schlosshauer:2019:qd}, implementation of a protective measurement will almost certainly have to include an analysis of the influence of the environment such as we have given here. 

We stress that the study presented in this paper is independent of any particular physical realization of the qubit system and the apparatus. It describes how environmental interactions affect a generic protective qubit measurement when such interactions cannot be avoided. What is perhaps surprising about our results is how significant the impact of the environment the measurement can be even when the decohering influence on the quantum state of the system is small. 


%

\end{document}